\documentclass[%
 reprint,
 superscriptaddress,
showpacs,preprintnumbers,
nofootinbib,
 amsmath,amssymb,
prl,
]{revtex4-1}

\usepackage{graphicx}
\usepackage{dcolumn}
\usepackage{bm}
\usepackage{color}

\begin{document}

\title{Detecting Lorentz Violations with Gravitational Waves from Black Hole Binaries}

\author{Thomas P.~Sotiriou}
\affiliation{School of Mathematical Sciences, University of Nottingham, University Park, Nottingham, NG7 2RD, UK}
\affiliation{School of Physics and Astronomy, University of Nottingham, University Park, Nottingham, NG7 2RD, UK} 

\date{\today}

\begin{abstract}
Gravitational wave observations have  been used to test Lorentz symmetry by looking for dispersive effects that are caused by higher order corrections to the dispersion relation. In this Letter I argue on  general grounds that, when such corrections are present, there will also be a scalar excitation. Hence, a smoking-gun observation of Lorentz symmetry breaking would be the direct detection of scalar waves that travel at a speed other than the speed of the standard gravitational wave polarisations or the speed of light. Interestingly, in known Lorentz-breaking gravity theories the difference between the speeds of scalar and tensor waves is virtually unconstrained, whereas the difference between the latter and the speed of light is already severely  constrained by the coincident detection of gravitational waves and gamma rays from a binary neutron star merger.
\end{abstract}

\maketitle

A deviation from general relativity may introduce changes in both the generation and the propagation of gravitational waves. Modeling the former is significantly harder, and, hence, it is appealing to first  derive constraints using the latter. Indeed, the recent binary black hole and neutron star observations have given new constraints on the dispersion relation of gravitational waves \cite{Abbott:2017vtc,TheLIGOScientific:2017qsa,Monitor:2017mdv}. LIGO uses the parametrisation $E^2=c^2 p^2+A c^\alpha p^\alpha$, where $c$ is the speed of light and $\alpha>0$ \cite{Mirshekari:2011yq}. It should be noted that $A$ has dimensions and $[A]=[E]^{2-\alpha}$.  Constraints on $A$ for $\alpha=0$ can be interpreted as constraints on the mass of the graviton.  Terms with $\alpha\geq 1$ are strongly associated with Lorentz symmetry violations in gravity and they will be our primary concern here.

It is noteworthy that deriving any accurate  measurement of $A$ for $\alpha=2$, {\em i.e.}~the speed of gravitational waves if there is no dispersion, would require knowledge of  the distance from the source and the time of travel. This explains why the known binary black hole detections have not yielded a constraint. It has been recently pointed out that one can actually use the time delay between signals in the two LIGO detectors to obtain an upper bound. Using the first binary black hole detection  the speed of gravitational waves, $c_g$, can be up to 1.7 times the speed of light \cite{Blas:2016qmn}. This is a very week bound that is based on the assumption that the maximum extra distance that the wave had to travel to reach the second detector is the distance from the first. Only an event whose location is almost collinear with the location of the two  detectors would give a noteworthy improvement.

Combining events in a Bayesian approach could allow one to get close to the  1\% level if a signal is seen by multiple detectors \cite{Cornish:2017jml}. This resembles the accuracy with which one can constrain $c_g$ in known Lorentz-violating theories using other gravitational experiments. Ho\v rava gravity \cite{Horava:2009uw,Sotiriou:2009gy,Sotiriou:2009bx,Blas:2009qj,Sotiriou:2010wn} offers an illuminating example: a combination of stability conditions, vacuum Cherenkov constraints, post-Newtonian constraints, and binary pulsar constraints forces $c_g$ to differ from the speed of light by no more than 1\% range at best. This is a straightforward but rough estimate based on the constraints on $\beta$ in Ref.~\cite{Yagi:2013ava} and the relation $c_g^2=(1-\beta)^{-1}$.

On the other hand, the recent observation of a neutron star merger with an electromagnetic counterpart implies that $c_g$ cannot differ from the speed of light by more that a part in $10^{15}$ \cite{TheLIGOScientific:2017qsa,Monitor:2017mdv}. This constraint is obtained by comparing the time of the merger as estimated by the gravitational waves signal with the time of arrival of gamma ray radiation. Clearly, it surpasses  any previous bound on $c_g$ by several orders of magnitude. The reason that I chose to discuss other constraints in some detail will become apparent later on.

Let us now return to the parametrisation of the dispersion relation and discuss the constraints on $A$ for different values of $\alpha$. 
Since $[A]=[E]^{2-\alpha}$, all corrections except the mass term and the  $\alpha=1$ term are suppressed by positive powers of $c p/M_\star$, where $M_\star$ is some characteristic energy scale. Gravitational waves have very long wavelengths compared to what  would correspond to realistic values for $M_\star$, and this suggests a very strong suppression. This effect is actually mitigated by the fact that dispersive effects change the phase of gravitational waves and this is cumulative, {\em i.e.}~it adds up with the distance traveled \cite{Mattingly:2005re,Mirshekari:2011yq,Kostelecky:2016kfm}. This significantly enhances the constraints one gets, but not enough to make them theoretically interesting for higher order corrections to the dispersion relation. For instance, for $\alpha=3$ the upper bound on $M_\star$ is in the $10^9GeV$ range but for $\alpha=4$ it drops well below the $eV$ range \cite{Abbott:2017vtc,Mirshekari:2011yq,Yunes:2016jcc}.

Hence, currently gravitational wave observations do place some (marginally) interesting constraints on the mass term and the linear and the cubic corrections to the dispersion relation of gravitational waves. However,  it is straightforward to exclude all of these terms in theoretical model building without restricting attention to a specific theory. The linear and cubic terms  can be excluded by imposing parity. The mass of the graviton can just be set to zero and it will be protected from quantum corrections by symmetry (in fact, it is actually hard to give the graviton a mass within the framework of a consistent theory, as is well known \cite{deRham:2014zqa}). 

This is rather disappointing when seen from the theory perspective: any model with parity invariance and no mass will have a dispersion of the type
\begin{equation}
\label{dispersion}
E^2= c_g^2 \,p^2+ \beta \frac{p^4}{M_\star^2}+ \gamma \frac{p^6}{M_\star^4}\ldots
\end{equation}
where I have set $c=1$, and $\beta$ and $\gamma$ are dimensionless. Given that $c_g$ is already so severely constrained, to have any hope of detecting or constraining further Lorentz violations from dispersive effects through gravitational wave observations one would have to have $M_\star$ of the order of a few $meV$. This is rather unlikely considering: (i) that the motivation for such terms in the dispersion relation comes from quantum gravity and (ii) such values for $M_\star$ are getting quite close to the characteristic length scales (sub-$mm$) at which we already test gravity in the lab. 

 The purpose of this Letter is to clarify that there is  a `smoking gun', direct detection of Lorentz symmetry breaking in gravity  that is not (directly) related to a precise measurement of $c_g$ and does not require measuring  dispersive effects. Hence, it does not require knowledge of the precise location of the source and the time of travel of the signal and it can be applied equally well to binaries with no electromagnetic counterpart at all. Moreover, it does not hinge on the value of the energy scale $M_\star$. All of the above can be argued without reference to any specific Lorentz-violating gravity theory.

The first step is to realise that a theory that leads to the dispersion relation  of eq.~\eqref{dispersion} will generically have a scalar degree of freedom as well. Eq.~\eqref{dispersion} is not invariant under Lorentz boosts, as the latter mix time and spatial derivatives. In fact, eq.~\eqref{dispersion} is derived via a Fourier transform of a differential equation that is second order in time derivatives and higher order in spatial derivatives. If one were to perform an arbitrary boost, this equation would end up being higher-order in time derivatives. That is, it is not enough to employ a flat-space approximation to drive this dispersion relation. One also needs to select a specific spacelike {\em foliation} of Minkowski space in order to render it physically meaningful. 

Let me now make the rather minimal assumption that the gravity theory from which eq.~\eqref{dispersion} is derived can admit a geometric description in terms of a metric $g_{\mu\nu}$. Local Lorentz symmetry can be roughly thought of as the local counterpart of diffeomorphism invariance, which is related to invariance under general coordinate transformations. The fact that linearised perturbations of the metric satisfy an equation that is second order in time derivatives and higher order in spatial derivatives, as implied by  eq.~\eqref{dispersion}, will persist around any background. This essentially follows by the local flatness theorem, which implies that the equation determining the evolution of metric perturbations should reduce to eq.~\eqref{dispersion} in a small enough patch. But as in the flat-space case above, any statement about the differential order in time derivatives is only valid in a specific foliation. If one were to perform a transformation of the type $t\to \tilde{t}(t, x^i)$, then the perturbation equation would become higher-order in time derivatives. 

The arguments above imply that the gravity theory that gives rise to a dispersion relation of the type of eq.~\eqref{dispersion} can only be sensible if it has a preferred foliation. Sensible in this context means ghost free and unitary. These technical requirements relate to stability and posi\-tivity of energy and allow the theory to be predictive in the appropriate sense both classically and quantum mechanically. If there is a preferred foliation, and if the field content (without matter) is just the metric, the maximal amount of symmetry one can have is diffeomorphisms that respect this foliation. This is a combination of time reparametrizations and 3-diffeomorphisms, {\em i.e.}~transformations of the kind $t\to\tilde{t}(t)$, $x^i\to \tilde{x}^i(t,x^i)$. This is less symmetry than general relativity has with the same field content, and, hence, there will be more dynamical degrees of freedom. 

One way to assess the number and type of extra degrees of freedom is to recall that symmetries generate constraints and counting these constraints allows one to calculate the number of degrees of freedom given the field content. Another, more intuitive way to arrive to the same result is to think geometrically. A foliation corresponds to the level surfaces of a scalar field, which I will call $T$. For the surfaces of the foliation to be spacelike, the gradient of $T$ should be timelike, as $u_\mu\equiv \nabla_\mu T/(\sqrt{g^{\alpha\beta}\nabla_\alpha T\nabla_\beta T})$ is the unit normal to the leaves of this foliation. One can now observe that, when given the pair $(g_{\mu\nu},T)$ in the context of a diffeomorphism invariant theory, one can always use part of the coordinate freedom to choose $T$ as a time coordinate. This would make $T$ trivial and it would reduce the symmetry to the foliation-preserving transformations discussed above. Alternatively, one can always start with the original theory that was invariant under foliation-preserving diffeomorphisms only and restore full diffeomorpism invariance by introducing a scalar field $T$ in an appropriate way. This symmetry restoration process is generally known as the Stueckelberg trick \cite{Ruegg:2003ps} (see also Refs.~\cite{Germani:2009yt,Jacobson:2010mx,Blas:2010hb} for direct applications in Lorentz-violating gravity) and it relates dynamically equivalent theories. The diffeomorphism invariant version of the theory makes the existence of an extra scalar degree of freedom explicit.

Two comments on the genericity of the arguments presented above are due. First, foliation-preserving diffeomorphisms is the maximal amount of symmetry that a theory that yields eq.~\eqref{dispersion} can have. Additionally, I assumed above that the theory had the minimal field content while having this symmetry --- just the metric.  Hence, the theory in question can always have more extra degrees of freedom that the scalar identified above.   Second, one could carefully design the theory so as to excise the scalar mode. Doing so via tuning parameters of a certain model appears particularly hard to achieve and is most likely to lead to naturalness issues from a quantum field theory perspective. A more appealing possibility is to make sure that the model has an extra symmetry. A characteristic example has been proposed in Ref.~\cite{Horava:2010zj}. It is worth stressing though that introducing the extra symmetry in this model requires adding auxiliary fields in a very specific way, and it is unclear if such a structure is stable under radiative corrections. I believe it is fair to say that such models are certainly not generic. 

From the discussion above one can conclude that a {\em generic} theory that leads to the dispersion relation of eq.~\eqref{dispersion}  will have {\em at least} one extra dynamical scalar degree of freedom. The requirement that this scalar mode be well-behaved dynamically in the infrared/large wavelength limit and in the linearised regime, implies that scalar excitations will satisfy a wave equation. In the context of Lorentz-violating theories with a dispersion relation \eqref{dispersion}, these excitations can be thought of as waves of the field  $T$ or as longitudinal modes of the metric. These two interpretations are equivalent and depend in the choice of foliation (see discussion above regarding the restoration of diffeomorphism invariance). Moreover, the fact that every solution should have a preferred foliation is equivalent to the requirement that $T$ is non-trivial in every spacetime, black hole spacetimes included. Hence, one expects that, in principle, this scalar mode will be excited in black hole binaries and lead to dipolar emission ({\em e.g.}~see Ref.~\cite{Barausse:2016eii}). (In Ho\v rava gravity, where black holes have been studied, they do have a non-zero charge associated with $T$, but this change is fully determined by the mass \cite{Eling:2006ec,Barausse:2011pu,Barausse:2013nwa}. Unfortunately, it is not yet worked out how this charge, and the relation that it has to satisfy with the mass, affect the efficiency with which scalar waves are emitted in black hole binaries.)

There is no reason to expect that the speed of these scalar waves, $c_s$, will be the same as $c_g$ above. Let me assume that one would detect such waves. Measuring $c_s$ would require knowledge of the distance from the source and the travel time, same as for measuring $c_g$. However, there are two other interesting possibilities. 

First, one could try to measure or constrain $c_s$ using the time delay between different detectors, as suggested in Ref.~\cite{Blas:2016qmn} for $c_g$. As discussed above, for this method to work one needs to have rather large deviations from the speed of light, $c$. Indeed,  I argued above that in known theories there are strong enough constraints on $c_g$ that make a detection of Lorentz symmetry breaking via this method quite unlikely. However, in certain theories  $c_s$ could have significantly larger deviations from $c$ than $c_g$ without violating viability conditions.   Ho\v rava gravity is such an example and it has been shown explicitly in Ref.~\cite{Gumrukcuoglu:2017ijh} that $c_s$ remains virtually unconstrained to date. Though this will not be necessarily true for every Lorentz-violating theory, it is  likely to be quite generic: observations push viable theories parametrically closer to general relativity. However, in that limit,  and in order to recover local Lorentz symmetry, the scalar mode has to either decouple or disappear. Hence, there is no reason to believe that  $c_s$ has to approach the speed of light. So, from a theory-agnostic perspective there is hope in directly detecting Lorentz symmetry breaking by the time delay  between detectors for scalar waves.

Second, one could base a detection simply on the time delay between the arrival of tensor and  scalar waves at the same detector. Assuming no dispersion, the difference in the time of arrival for two waves traveling at speeds $c_1$ and $c_2$ is 
\begin{equation}
\label{Dt}
\Delta t = \frac{d}{c_1} \left(\frac{c_1}{c_2} -1\right)= t_1 \left(\frac{c_1}{c_2} -1\right),
\end{equation}
where $d$ is the distance traveled and $t_1$ the travel time for the wave with speed $c_1$. It should be clear from eq.~\eqref{Dt} that even extremely small differences between the speeds will give a very significant time delay, as the sources are very distant and travel times are very long. This method would be particularly efficient in detecting  deviation from Lorentz symmetry if there are any. Said otherwise, measuring a nonvanishing $\Delta t$ is a smoking gun for Lorentz symmetry breaking.

The two approaches for constraining $c_s$ discussed above are complementary. Using the time delay between detectors will fail if $c_s$ does not differ significantly from the speed of light. On the other hand, using the time delay between the usual gravitational wave polarizations and scalar waves can only work in practice if the difference between $c_s$ and $c_g$ is extremely small, otherwise the difference in arrival time would simply be too large and it would be impossible to associate the signals. For example,  eq.~\eqref{Dt} clearly implies that a 1\% fractional deviation in speeds between 2 signals would lead to a $\Delta t$ that can be of the order of millions of years for gravitational wave sources. To get $\Delta t$ less than a year, in order to have both signals within the same observation run of a network of detectors, would require a fractional deviation in speeds of a few parts in a billion. 

I have argued above why Lorentz breaking gravity theories that exhibit dispersive effects will also generically have a scalar excitation. I have also pointed out that detecting scalar waves that travel at a speed that is different from  the speed of the standard gravitational wave polarisations or the speed of light would be a smoking gun detection for Lorentz symmetry breaking. These statements and the supporting arguments might not surprise to experts  in Lorentz-violating theories. The purpose of this Letter if to make them more accessible to the broader gravitational physics community and to motivate further study on how to detect new gravitational wave polarisations and determine their speed. 

Clearly, there are very significant challenges related to this suggestion. The most obvious one is related to  detectability.  Current gravitational wave detectors are not optimised for picking up scalar waves. A detection would require multiple detectors \cite{Chatziioannou:2012rf} but also an appropriate signal. It might  well be that scalar emission is severely suppressed throughout the binary evolution. For instance, in standard (Lorentz invariant) scalar-tensor theories, weak field test and binary pulsar observations already lead to severe indirect constraints on dipolar emission  from neutron star binaries, see for example  Refs.~\cite{Damour:1998jk,Barausse:2012da}). Additionally, in certain theories it might be more likely to detect the existence of a longitudinal mode indirectly, by considering how dipolar emission affects the binary dynamics and how this, in turn, affects the phase of the standard polarisations \cite{Yunes:2016jcc}. However, these statements are both theory and system dependent.
 
Another significant challenge is associating signals with each other and attributing them to the same event when deviations from Lorentz symmetry are significant enough to lead to appreciable differences in speeds. As I discussed above, for known Lorentz-breaking gravity theories, such as Ho\v rava gravity, once can easily satisfy all known constraints and still have propagations speeds that differ enough to lead to arrival times separated by years or even millions of years for distant events.  

The extreme sensitivity of the difference in arrival times on differences in speeds certainly offers exciting opportunities. This is already demonstrated in practice by the bound on $c_g$ that was obtained by the neutron star merger with an electromagnetic counterpart \cite{Monitor:2017mdv}. 
It is however also  a significant limitation when it comes to detecting Lorentz symmetry breaking with gravitational waves and its role has not been stressed enough. It implies that detecting a longitudinal mode that travels at a speed that is different than that of the usual gravitational wave polarisation might require examining very long stretches of data and going beyond standard match filtering techniques. This question certainly deserves further investigation.

Despite the aforementioned challenges, searching for longitudinal gravitational waves that travel at a different speed comes with a  significant payoff: as argued above, it can offer direct, independent constraints on Lorentz symmetry violations. Hence, it is a possibility which should be explore further, in parallel with all other methods.

I was motivated to write this Letter after chairing a discussion session at the workshop ``New Frontiers in Gravitational-Wave Astrophysics'' that took place in Rome. I would like to thank the organisers and all of the participants of the workshop. I am indebted to Enrico Barausse and Nico Yunes for a critical reading of an earlier version of this manuscript and for providing numerous enlightening comments.
The research leading to these results has received funding from the European
Research Council under the European Union's Seventh Framework Programme
(FP7/2007-2013) / ERC grant agreement No.~306425 ``Challenging General
Relativity''. I also acknowledge financial support from STFC consolidated grant No.~ST/P000703/1 and networking support by the COST Action GWverse CA16104.


\end{document}